# Application of boundary functionals of the theory of random processes to aerosol coagulation

V. V. Ryazanov

*Institute for Nuclear Research, pr. Nauki, 47 Kiev, Ukraine, e-mail: vryazan19@gmail.com*

**Abstract**

A new approach to describing aerosol behavior is proposed. Boundary functionals of random process theory are applied to describe the behavior of aerosol concentrations during coagulation. It is shown that considering the first-passage time of a given aerosol concentration level corresponds to experimental results for the time dependence of aerosol concentration. Probabilities for aerosol concentrations to attain specific values are obtained, as well as expressions for average aerosol concentrations.

## 1. Introduction

Stochastic models can describe the evolution of aerosol systems, including the number and size of clusters, as they combine over time. For example, a birth-and-death formalism can be used to model the number of aerosol clusters. So, the birth and-death model leads to the kinetic equation of coagulation in the form

$$\partial P(t, [X])/\partial t = (1/2)\sum_{i \neq j} W(i,j) \left[ (X_i+1)(X_j+1) P(X_i+1, X_j+1, X_{i+j}-1) - X_i X_j P(t, [X]) \right] +$$
$$+ (1/2)\sum_i W(i,i) \left[ (X_i+2)(X_i+1) P(X_i+2, X_{2i}-1) - X_i(X_i-1) P(t, [X]) \right], \qquad (1)$$

where $P(t, X_1, X_2, ..., X_n, ...)$ is the probability to find $X_i$ particles (clusters) having the size $i$ ($i=1,2,...$) in the time $t$; $W(i,j)$ is the coagulation probability per time unit of the particles $i$ and $j$ (containing, in general, the factor $L^{-3}$, where $L$ is the size of a system). From equation (1) in (Lushnikov, 1978; Ryazanov, 2011), the Smoluchowski coagulation equation was obtained.

The stochastic description of the coagulation process was performed in papers like (Scott, 1967; Bayewitz et all, 1971; Williams, 1984; Merculovich & Stepanov, 1985, 1991, 1992; Van Dongen & Ernst (1987); Ryazanov, 1991, 2011; Kolodko, Sabelfeld & Wagner, 1999; Lushnikov, 2003; Bhatt & Ford, 2003; Debry, Sportisse & Jourdain, 2003) and other. Many stochastic studies of aerosol systems are heuristic in nature. For example, in the works (Merculovich & Stepanov, 1985, 1991, 1992), a distribution was obtained that was later "rediscovered" by mathematicians.

In the theory of random processes, boundary functionals associated with the process reaching its boundaries (i.e., functionals from selective trajectories associated with the exit of these trajectories beyond certain boundaries) occupy an important place. These include the first-passage time (*FPT*), process extrema, moments of reaching extrema, time spent above the level, moments of process return to a certain region, and other functionals with physical meaning. Thus, the first-passage time is widely used in the study of various physical, chemical, biological, and economic problems (e.g., (Metzler et al, 2014)). Examples of the application of other boundary functionals can be found in (Ryazanov, 2025a). Aerosol coagulation is described by equations with complex and cumbersome solutions (Voloschuk, 1984; Piskunov, 2010). Additional hypotheses are put forward, such as the possibility of an anomalously large particle, not described by the Smoluchowski equation, whose size immediately goes to infinity and whose concentration is infinitesimally small. This article provides a general outline of the potential application of boundary functionals to the description of aerosol behavior. Simple stochastic coagulation models (e.g., a constant coagulation constant) are used. However, agreement with the results of (Fuchs, 1964) is obtained. Only two examples of boundary functionals are considered, of which there are many more. This is primarily due to the behavioral characteristics of the quantity under consideration, the aerosol concentration, and its decrease over time. This analysis represents an original, previously unused approach to describing aerosols, and is independent of many papers on aerosol coagulation kinetics, such as those cited in (Voloschuk, 1984; Piskunov, 2010).

Section 2 introduces the stochastic model used in the article. Section 3 characterizes the boundary functionals. Section 4 demonstrates the application of boundary functionals to the description of aerosols. Section 5 concludes the article.

## 2. Stochastic storage model and the Smoluchowski equation

In this article, random processes for the aerosol concentration $n(t)$ and the ratio $\xi(t)=n(t)/n_0$, $n_0=n(t=0)$ are considered. It is also possible to consider other characteristics of aerosols: their size, coagulation coefficient, fractal properties, etc. Since aerosol particles combine during coagulation, the particle concentration $n(t)$ and $\xi(t)$ are non-increasing processes. In the article (Ryazanov, 2025b), a non-decreasing random process for the activity $K(t)$ was considered. Here, the situation is the opposite. In the terminology of (Gikhman & Skorokhod, 1969; Gusak, 2007), such processes are called semi-continuous from above and below. In this case, restrictions are imposed on the boundary functionals. Thus, for $n(t)$ and $\xi(t)$ there are no return moments and time spent above the level: the moments $\tau'(u) = \inf\{t > \tau(u), R(t) > 0\}$, which is the moments of return of the process $R(t)$ after bankruptcy ($R(t) \leq 0$) to the half-plane $\Pi^+ = \{y > 0\}$, where $R(t)$ is the reserve risk process, which takes the form: $R(t) = u + ct - S(t)$, where u is initial value, $c>0$ is constant premium rate, $S(t) = \sum_{k=1}^{N_t} \xi_k$, $P\{\xi_k \neq 0\} = 1$, represent the amounts of payments (claims), $\xi_k > 0$ in the risk theory (Rolsky, 1999; Asmussen, 2000; Gusak, 2007); time spent above the level coincides with *FPT*, etc.

The stochastic storage process (Cinlar & Pinsky, 1972; Brockwell & Chung, 1975; Çinlar, 1975; Harrison & Resnick, 1978; Rubinovitch & Cohen, 1980; Brockwell, Resnick & Tweedie, 1982; Zakusilo, 1988, 1990) is described by the equation

$$\frac{dX(t)}{dt} = \frac{dA}{dt} - r_\chi[X(t)], \qquad (2)$$

where $X(t)$ is a random variable of the stock in the system, $dA(t)/dt$ is a random rate of stock entry into the system, $r_\chi[X(t)]$ is the exit rate. The exit rate takes into account the singularity at zero: $r_\chi[X(t)] = r(X(T)) - r(0+)\chi_{X(t)}$; $\chi_X = 1$, at $X = 0$, . This is due to the fact that the storage model (2) is defined for non-negative values of $X(t) \geq 0$, and the exit from an empty system must be zero. The function $r[X(t)]$ can be chosen arbitrarily. The entry function $A(t)$ can be described by various classes of random processes. In (Cinlar & Pinsky, 1972; Çinlar, 1975), the entry rate is characterized by a Levy process with non-decreasing trajectories and zero drift. Brownian motion and the Poisson process also belong to Levy processes. Stochastic storage processes were applied to the description of aerosol coagulation in (Ryazanov, 1991, 2011). The Laplace transform $E(\exp\{-\theta A(t)\})$ of the entry function $A(t)$ is associated with the function $\varphi(\theta)$, the so-called cumulant (or *SCGF* (11)) of a process $A(t)$ of the form:

$$E(\exp\{-\theta A(t)\}) = \exp\{-t\varphi(\theta)\} = \int_0^\infty e^{-\theta x} k(x,t)dx; \quad \varphi(\theta) = \int_0^\infty (1-\exp\{-\theta x\})\lambda g(x)dx, \quad (3)$$

$$\lambda = \varphi(\infty) < \infty; \quad \rho \equiv \lambda \int xg(x)dx = E(\frac{dA}{dt}) = \frac{d\varphi(\theta)}{d\theta}\Big|_{\theta=0}; \quad \mu^{-1} \equiv \int xg(x)dx = \frac{1}{\varphi(\infty)}\frac{d\varphi(\theta)}{d\theta}\Big|_{\theta=0},$$

where $E(...)$ denotes averaging, $k(x,t) = P(x \leq A(t) \leq x+dx)$ is the probability density function of the random variable $A(t)$, $\lambda$ is the intensity of the input flow, and the density function $g(y)$ describes the magnitude of the jumps in the inflows. In representation (3), the input process corresponds to a jump-like generalized Poisson process, where inflows occur at random moments in time in randomly sized portions. If we use the analogy with a reservoir and the water supply in it, then the quantity $\lambda$ describes the intensity of Poisson random jumps, moments in time when water inflows into the system occur. The density function $g(y) = P(y \leq m \leq y+dy)$ describes the random amount of water $m$ entering the reservoir in one jump, with an average of $\mu^{-1}$.

Another approach to the stochastic storage model was proposed by Zakusilo (1990). This approach is based on the observation that processes of type (2) fit into the general scheme of dynamic systems subject to the additive effect of process A(t). This definition and some properties of the storage process are given by Zakusilo (1990). Truncated storage process is considered, taking values from a certain compact set [0, *a*], and the average times for reaching the lower level by inventory storage processes are investigated.

The approach proposed by Zakusilo (1990) extends the applicability of storage processes. It directly links storage processes to dynamic systems and replaces the most commonly used random disturbances in the form of white noise with more general generalized Poisson processes.

Let us define the function $g(y)$ in explicit form of an exponential distribution:

$$g(x) = \mu e^{-\mu x}, \quad \mu > 0, \quad \varphi(\theta) = \lambda\theta/(\mu+\theta). \quad (4)$$

If $\mu = 1$, then $\rho = \lambda$.

Let's compare the stochastic storage model (2) with the dynamics of aerosol coagulation. Consider a simplified model of the Smoluchowski kinetic equation of coagulation for an aerosol concentration *n(m,t)* of mass *m* of the form:

$$\frac{dn(m,t)}{dt} = \frac{1}{2}\int_0^\infty K(m, m-m_1)n(m_1)n(m-m_1)dm_1 - n(m)\int_0^\infty K(m,m_1)n(m_1)dm_1, \quad (5)$$

where *m* is the mass of the aerosol particle, *K* is the coagulation constant. We use the expression for the asymptotic solution at *K=const* (Schumann, 1940), in which

$$n(m) = \frac{ce^{-m/\bar{m}}}{(\bar{m})^2}, \quad \bar{m} = \frac{c}{n} = \frac{c}{n_0}(1+0.5Kn_0 t) \approx \frac{Kct}{2}. \quad (6)$$

Equation (5) is comparable with the storage model equation (2) with the output function *r(n)=bn, n(t)=X(t)*.

If we consider equation (5) as an equation for the average value of concentration, then at *K=const* it can be compared with the average value of equation (2) and substituting (6) into (5) at $c \approx n_0$ it can be written in the form:

$$\frac{dn(m,t)}{dt} = \rho - bn(m,t), \quad \rho = \frac{1}{2}\int_0^\infty K(m, m-m_1)n(m_1)n(m-m_1)dm_1 = \frac{1}{2}K\frac{c^2 e^{-m/\bar{m}}}{(\bar{m})^4}m, \quad (7)$$

$$b = \int_0^\infty K(m,m_1)n(m_1)dm_1 = Kc/\bar{m} = Kn = Kn_0/(1+0.5Kn_0 t), \quad \bar{m} = \frac{c}{n}.$$

In the storage model (2), the parameters $\rho$ and $b$ are independent of time. The time dependence in (7) can be eliminated by, for example, setting $t$ to some average value $<t>$. Or by averaging over time in the interval $T-t_0$. Then

$$\bar{b} = \int_{t_0}^T b\,dt/(T-t_0) = 2\ln\frac{(1+0.5Kn_0 T)}{(1+0.5Kn_0 t_0)}/(T-t_0),$$

$$\bar{\rho} = \frac{m}{(T-t_0)}[e^{-m/(1+0.5Kn_0^2 T)}(\frac{1}{m(1+0.5Kn_0 T)^2}+\frac{2}{m^2(1+0.5Kn_0 T)}+\frac{2}{m^3}) -$$

$$e^{-m/(1+0.5Kn_0 t_0)}(\frac{1}{m(1+0.5Kn_0 t_0)^2}+\frac{2}{m^2(1+0.5Kn_0 t_0)}+\frac{2}{m^3})]. \qquad (8)$$

For storage process (2) with output function $r(n)=bn$ the Laplace transform of the solution Eq. (7), random variable $X(t)=B(t)$ from (2), (3) has the form:

$$F_{bq}(e^{-\theta},t) = \mathbf{E}(e^{-\theta B(t)}|B(0)=B_0) = \exp\{-\theta B_0 e^{-bt} - \lambda\int_0^t[1-\Psi(\theta e^{-bu})]du\}, \quad \Psi(\theta) = \int_0^\infty \exp\{-\theta x\}g(x)dx. \qquad (9)$$

For the Laplace transform of the stationary probability density of the random variable $X$, we obtain that:

$$\ln F(\exp\{x\}) = -\int_0^{-x}(\varphi(u)/u)du/b; \quad A(x) = \langle B(x)\rangle = \varphi(-x)/b(-x);$$

$$f(A) = \rho - bA(x) = dA/dt; \quad A(t) = \exp\{-bt\}(A_0 - A_{st}) + A_{st}; \quad A_{st} = \rho/b. \qquad (10)$$

## 3. Boundary functionals

The characteristic function of a homogeneous process $\xi(t)$, $t \geq 0$ is determined in the theory of random processes (for $\xi(0)=0$) (Feller, 1957; Borovkov, 1976; Gikhman & Skorokhod, 1969; Gusak, 2007) by the relation:

$$\mathbf{E}e^{i\alpha\xi(t)} := \int_{-\infty}^\infty e^{i\alpha x}dF(x) = e^{t\Psi(\alpha)}, \quad t \geq 0, \qquad (11)$$

where $F(x) = P(\xi < x)$ is the distribution function of a random process $\xi(t)$, $t \geq 0$, and the function $\psi(\alpha)$ represents the scaled cumulant generating function (*SCGF*) (or simply cumulant) of the process $\xi(t)$, $t \geq 0$. The *SCGF* do not depend on $t$ and characterize all finite-dimensional distributions of the process.

If for a process $\xi(t)$, $t \geq 0$, the function $\psi(\alpha)$ at $i\alpha = r$ is equal to $s$, then we obtain the equation:

$$\Psi(\alpha)|_{i\alpha=r} := k(r) = s, \quad \pm\text{Re}\,r \geq 0. \qquad (12)$$

This equation, in risk theory (Rolsky et al., 1999; Asmussen, 2000; Gusak, 2007) is referred to as the fundamental Lundberg equation. This equation has two roots; $r_1 = -\rho_-(s)$, $r_2(s) = \rho_+(s) > 0$. Knowledge of roots $\rho_-(s)$, $\rho_+(s)$ is important for the study of boundary functionals.

3.1. The forms of boundary functionals

For certain functionals of the random process $\xi(t)$, definitions are provided in Refs. (Rolsky et al., 1999; Gusak, 2007). These include:

$$\tau(x) = \inf\{t : \xi(t) > x\}, \quad x > 0 \text{ is the moment of the first exit for the level } x; \qquad (13)$$

$\gamma(x) = \xi(\tau(x)) - x$ is first overjump over $x$;

$\zeta^{\pm} = \sup_{0 \le t < \infty} (\inf) \zeta(t)$ represents the absolute extremes of the process $\zeta(t)$, etc.

In risk theory, $\tau(u)$ is defined as the ruin time. In physics, this is called the first-passage time (*FPT*) (Metzler, Oshanin & Redner, 2014). We can consider the moment when the value of *ξ(t)* reaches zero, which represents the moment of degeneracy or the end of the period of existence. This is because in physics problems, the reaching of zero by a given process is often referred to as the end of the period of existence, of the lifetime of the quantity described by this process, or the *FPT* of the zero level.

### 3.2. The expressions for *FPT*

In the works (Borovkov, 1976; Rolsky et al., 1999; Gusak, 2007) an expression for the moment generating function (*MGF*) of *FPT* $\tau(x)$ (13) of the form $T(s,x) = \boldsymbol{E}[e^{-s\tau(x)}, \tau(x) < \infty]$ was obtained:

$$T(s,x) = \boldsymbol{E}[e^{-s\tau(x)}, \tau(x) < \infty] = e^{-\rho_+(s)x}, \quad x > 0, \quad (14)$$

$$\boldsymbol{E}[\tau(x), s] = -\frac{\partial \ln \boldsymbol{E}[e^{-s\tau(x)}, \tau(x) < \infty]}{\partial s} = \frac{\boldsymbol{E}[\tau(x)e^{-s\tau(x)}, \tau(x) < \infty]}{\boldsymbol{E}[e^{-s\tau(x)}, \tau(x) < \infty]} = x \frac{\partial \rho_+(s)}{\partial s}.$$

Exact solutions for the first-passage time of the storage model (2) are found only for a small number of cases. One of them is considered in the article by Zakusilo (1988), when for process (2) $r(x)=bx$, and $v(x,\infty)=\int_0^\infty \lambda g(x)dx = \lambda e^{-\mu x}$, $\lambda>0$, $\mu>0$, $v>0$, Eqs. (3), (4), (7). However, solutions of linear equations of the form $dX(t)=-bX(t)dt+dA(t)$ describe a whole series of real processes. If $v(x,\infty)=\lambda e^{-\mu y}$ for $x>0$, then:

$$\varphi(s;x,y) = F(\frac{s}{b}, \frac{s+\lambda}{b}+1, x\mu) \int_{x\mu}^{\infty} \frac{dz}{F^2(\frac{s}{b}, \frac{s+\lambda}{b}+1, z)p(z)} \Big/ F(\frac{s}{b}, \frac{s+\lambda}{b}+1, y\mu) \int_{y\mu}^{\infty} \frac{dz}{F^2(\frac{s}{b}, \frac{s+\lambda}{b}+1, z)p(z)}$$

$$p(x) = e^{-x} x^{\frac{s+\lambda}{\mu}+1}. \qquad \varphi(s;x,y) = \boldsymbol{E}_x e^{-s\tau(y)} \qquad (15)$$

Here *F(a,b,x)* is the Kummer function, which has the representation (Abramowitz & Stegun, 1972):

$$F(a,b,x) = 1 + \sum_{k=1}^{\infty} \frac{a(a+1)...(a+k-1)}{b(b+1)...(b+k-1)} \frac{x^k}{k!},$$

$x$ is the initial value of the random process; $y$ is the level of achievement.

If expression (15) is compared with formula (14), then the dependence $\rho_+(s)$ can be determined. Figure 1 shows a comparison of the functions $\rho_+(s)$ obtained in this way from equations (15), (14), Fig. 1a, (in our case $\rho_+(s=0)=0$ and $T(s=0,x) = \boldsymbol{E}[\tau(x) < \infty] = 1$) and from the Lundberg equation (12), where $k(r)$ is used as the function $\ln F(\exp\{x=r\})$ from expression (10), Fig. 1b. It is evident that the behavior of these functions is qualitatively identical, but quantitatively different. The results obtained for the mean values from expressions (14) and (15) also differ. This is due to the fact that expression (14) involves a conditional probability given $\tau(x) < \infty$, while in expression (15) the mean value is determined given the initial value $x$ of the random process. Expression (15) is not used below.

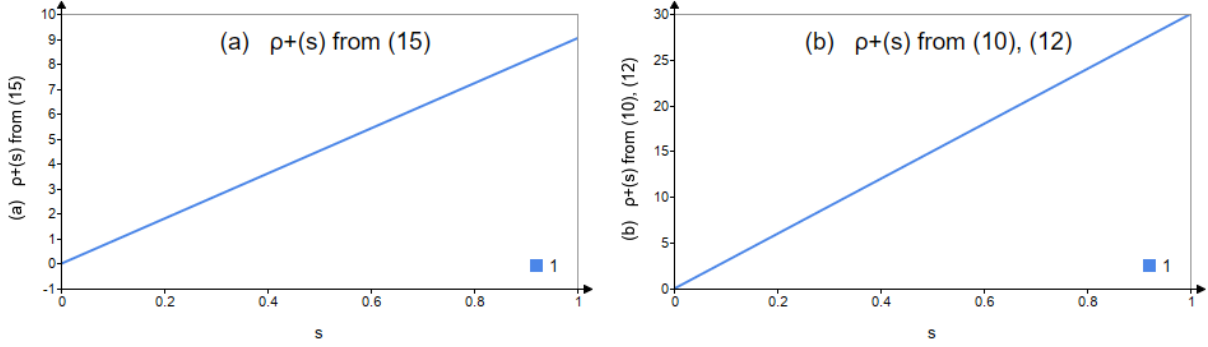

Fig. 1. Comparison of functions $\rho_+(s)$ obtained from relations (15) and (14), Fig. 1a, with $\rho_+(s)$ the obtained solution of the Lundberg equation (12) with the cumulant $k(r)$ from expression (10).

### 3.3. The solution of the Lundberg equation

As a function $k(r)$ in (12), taking into account expression (11) at $t=1$, we use expressions of the form $\ln F(\exp\{x=r\})$ from relations (9) and (10). In order for expression (9) to correspond to definition (11), we average it over time by performing the operation $\ln \bar{F}(r) = \int_{t_0}^{T} \ln F(r,t) dt / (T-t_0)$, as in (8). We obtain

$$\ln \bar{F}(r) = k(r) = re^{-\bar{b}T} + \frac{\bar{\lambda}}{\bar{b}} \ln \frac{1 - re^{-\bar{b}T}/\mu}{1 - r/\mu}. \tag{16}$$

For $T \to \infty$ we have the stationary case (10). For (10) the solution of Lundberg equation (12) is:

$$\rho_+(s) = \mu(1 - e^{-s\bar{b}/\bar{\lambda}}). \tag{17}$$

There is no negative root in this case. It can be obtained from expression (16) by expanding the second term on the right-hand side of (16) in a series, restricting ourselves to the quadratic term, and solving the quadratic equation:

$$\rho_+(s) = b_3[\sqrt{1 + s 2\mu^2 \bar{b}/\bar{\lambda} b^2_3} - 1] \approx s\mu^2 \bar{b}/\bar{\lambda} b^2_3, \quad \rho_-(s) = b_3[\sqrt{1 + s 2\mu^2 \bar{b}/\bar{\lambda} b^2_3} + 1], \tag{18}$$

$s>0$, $b_3 = e^{-\bar{b}T} \mu(n_0 \bar{b}\mu/\bar{\lambda} + 1/\bar{b}T)$.

## 4. Boundary functionals in aerosol theory

### 4.1. First-passage time

As in (13), we denote $\tau(y)$ the moment of the first achievement of the process $\xi(t)$ of the level $y=n_y(t)/n_0$ or $n_y(t)$. From (14) and (17) we obtain

$$<\tau(y,s)> = -\partial \ln E[e^{-s\tau(y)}]/\partial s = (\mu\bar{b}/\bar{\lambda})e^{-s\bar{b}/\bar{\lambda}} y. \tag{19}$$

Dependences on $s$ in the range $s=0,...,0.05$ at $y=0.5$ and on $y$ in the range $y=0,...,1$ at $s=0$ are shown in Fig. 2 a,b. We consider the process for $\xi(t)=n_y(t)/n_0$. The parameters used in the calculations are $m=30$, $\bar{m}(t=1160)=30$, $\bar{m}(T=10^5)=2501$, $n_0=10^8$, $T=10^5$, $\mu=25.8$, $\bar{\rho}=2.22\cdot 10^{-3}/T$, $\bar{b}=7.8/T$, $\bar{b}/\bar{\lambda}=135.5$, $\rho_+(s)=28.5(1-e^{-135.5s})$, $<\tau^+(y,s=0)>=3495.9y$, where $\bar{b}$, $\bar{\lambda}=\bar{\rho}\mu$ given in (7), (8).

We use on Fig. 2c relation (14), where the expression for $\rho_+(s)$ of the form (17) with parameters $m=18.6$, $\bar{m}(t=704)=18.6$, $\bar{m}(T=10^5)=2501$, $n_0=10^8$, $T=10^5$, $\mu=10$,

$\bar{\rho} = 5.73 \cdot 10^{-3}/T$, $\bar{b} = 7.8/T$, $\bar{b}/\bar{\lambda} = 135.5$, $\rho_+(s) = 10(1-e^{-135.5s})$ (compared to Fig. 2 a,b parameter µ changed from 25.8 to 10, parameter *m* - from 30 to 18.6). Then, $<\tau^+(y, s=0)>= 1355y$. In Fig. 2c shows the behavior of the quantity $10^6/y$, $y = n(t)$, from $<\tau(10^6/y, s=0)>$ where $<\tau(10^6/y, s=0)>$ is the time of first reaching the level $10^6/y$, corresponds to time t. Let's compare this figure with Fig. 71 for the kinetics of thermal coagulation from the book (Fuchs, 1964) (Fig. 2 d). Time to reach the level $10^6/y$, $y = n(t)$, the value $<\tau^+(10^6/y, s=0)>$ corresponds to the usual coagulation time t, which is plotted along the abscissa axis in Fig. 71 (Fuchs, 1964) (Fig. 2 d).

In this case we obtain Fig. 2c. This Fig. 2c almost completely coincides with Fig. 71 from the book (Fuchs, 1964), Fig. 2 d (the angle of inclination is slightly different; in Fig. 2c the time is in seconds, and 6000 sec = 100 minutes).

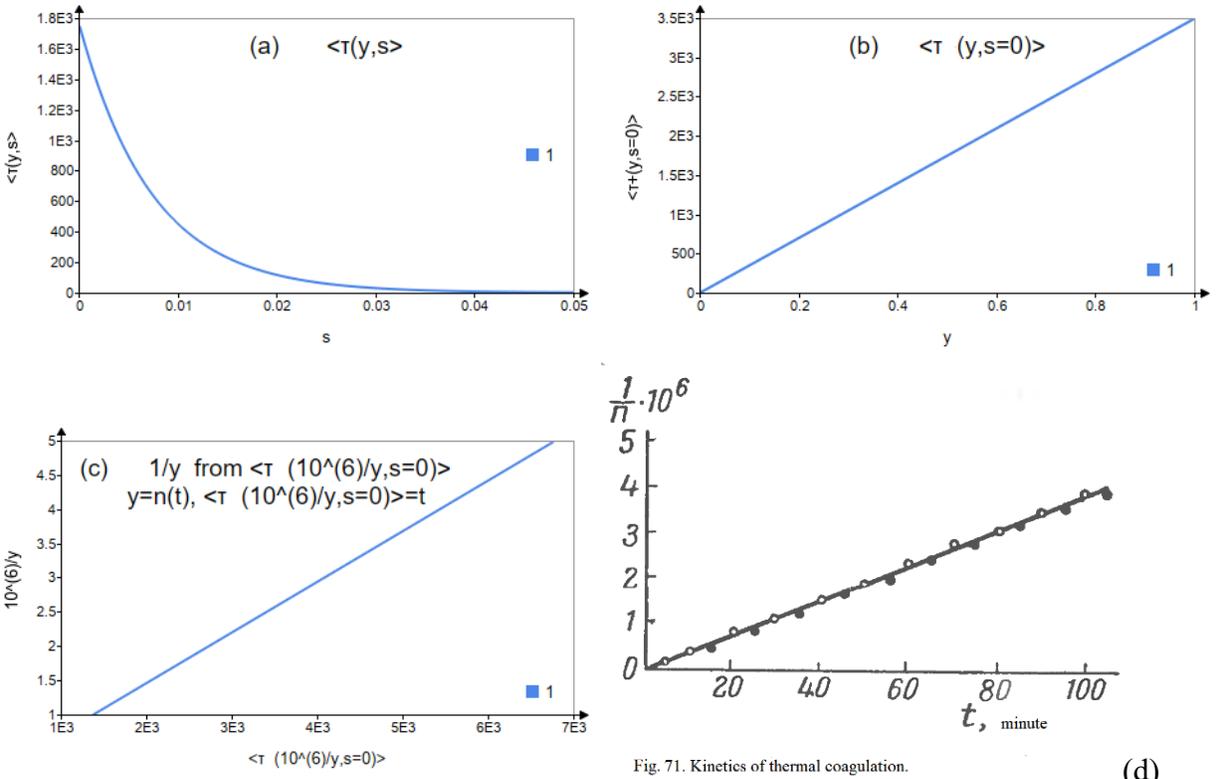

Fig. 2. Dependences of average values of FPT $<\tau(y,s)>$ from expressions (17), (19) on *s* in s=0,…,0.05 at y=n(t)/n₀=0.5, Fig. 2a, and on y=n(t)/n₀ in y=0,…,1 at s=0, Fig. 2b; $m = 30$, $\bar{m}(t=1160) = 30$, $\bar{m}(T=10^5) = 2501$, $n_0 = 10^8$, $T = 10^5$, $\mu = 25.8$, $\bar{\rho} = 2.22 \cdot 10^{-3}/T$, $\bar{b} = 7.8/T$, $\bar{b}/\bar{\lambda} = 135.5$, $\rho_+(s) = 28.5(1-e^{-135.5s})$, $<\tau^+(y, s=0)>= 3495.9y$. The behavior of the dependence $10^6/y$, $y = n(t)$ (but not n(t)/n₀) on $<\tau(10^6/y, s=0)>$, obtained from expressions (14), (17) at s=0, is shown in Fig. 2c; $m = 18.6$, $\bar{m}(t=704) = 18.6$, $\bar{m}(T=10^5) = 2501$, $n_0 = 10^8$, $T = 10^5$, $\mu = 10$, $\bar{\rho} = 5.73 \cdot 10^{-3}/T$, $\bar{b} = 7.8/T$, $\bar{b}/\bar{\lambda} = 135.5$, $\rho_+(s) = 10(1-e^{-135.5s})$; Fig. 2d copied from Fig. 71 (Fuchs, 1964).

In Fig. 2b, the increase in value $<\tau(y, s=0)>$ coincides with a decrease in time *t*. And in Fig. 2c, the increase in the average value $<\tau(10^6/y, s=0)>$ coincides with an increase in time *t*. The relationship between the coagulation time *t* and the random time τ(x) (Gikhman & Skorokhod, 1969) to reach level *x* depends significantly on *x*.

Such dependencies were also obtained from some other relationships for *FPT*. A comparison was made with the results of using expression (18) in (14). This comparison showed that approximation (18) very inaccurately describes the kinetics of thermal coagulation compared to approximation (17). Similar dependencies were obtained from expression (15), when $\tau(y,s) = -\partial \ln \varphi(s,x,y)/\partial s$.

Relationships of the type shown in Fig. 2c can be obtained from expressions (6). The use of *FPT* demonstrates the applicability of boundary functionals. This is used in the next subsection. The dependence of the average values of $FPT <\tau(y,s)>$ from expressions (17), (19) on s, shown in Fig. 2a, can be interpreted as a dependence on some physical field acting on the system.

4.2. Probability that the aerosol concentration takes a given value

For almost upper (lower) semi-continuous integer Poisson processes, the relation was obtained in (Gusak, 2007)

$$P\{\xi(t) = r\} = \frac{t}{r}\frac{\partial}{\partial t}P\{\tau(r) < t/\tau(0) < t\}, \quad r > 0, \tag{20}$$

where $P\{\xi(t) = r\}$ is the probability that the process $\xi(t)$ takes the value $r$, $P\{\tau(r) < t/\tau(0) < t\}$ is the probability that the first-passage time it takes for the process $\xi(t)$ to the level $r$ is less than the time t, provided that this is true for reaching the zero level. We assume that events $\{\tau(r) < t\}$ and $\{\tau(0) < t\}$ are independent. Then $P\{\tau(r) < t/\tau(0) < t\} = P\{\tau(r) < t\}$. The processes $\xi(t) = n(t)$ and $\xi(t) = n(t)/n_0$ are continuous from below and can be represented as Poisson processes (Scott, 1967; Ryazanov & Shpyrko, 1995). We use expression (19). We consider the process $\xi(t) = n(t)/n_0$. We determine the probability density $\frac{\partial}{\partial t}P\{\tau(r) < t/\tau(0) < t\}$ from the inverse Laplace transform of quantity (14). We proceed from the stationary Laplace transform (10) and a function $\rho_+(s)$ of the form (17). We choose the following parameter values: $m = 50$, $n_0 = 10^8$, T=$10^4$, $\bar{\lambda} = 0.33 \cdot 10^{-7}$, $\mu = 10$, $\bar{\rho} = 3.3 \cdot 10^{-9}$, $\bar{b} = 0.397 \cdot 10^{-3}$. Then $k(r) = -0.83 \cdot 10^{-4} \ln(1 - r/10)$, $\rho_+(s) = 10(1 - e^{-s \cdot 1.2 \cdot 10^4})$. For such a quantity $\rho_+(s)$, the inverse Laplace transform of expression (14) is not written explicitly. Function (17) $\rho_+(s) = 10(1 - e^{-s \cdot 1.2 \cdot 10^4})$ behaves as follows: at $s \geq 10^{-3}$ (which roughly corresponds to $t \leq 10^3$ sec) $\rho_+(s) = 10$. The transition region from $s$=0 to $s$=$10^{-3}$ is approximated by the dependence $e^{\alpha^{1/2}s^{1/2}}$, $\alpha^{1/2} = 154x$, x=n(t)/n₀ is the achievement level from expression (14). Note that other, more accurate approximations are also possible. Such approximations cannot be solved analytically, but can be calculated numerically.

The inverse Laplace transform of a function $e^{-x154s^{1/2}}$ has the form

$$\frac{154}{2\pi\sqrt{t^3}}xe^{-x^2 154^2/4t}. \tag{21}$$

Moving from a discrete probability distribution to a continuous one, from expressions (20)-(21) we obtain that

$$P\{\xi(t) = x\} = \frac{154}{2\pi\sqrt{t}}e^{-x^2 154^2/4t}, \tag{22}$$

$$\langle\xi(t)\rangle = \int_0^1 xP\{\xi(t) = x\}dx = \frac{154\sqrt{t}}{2\pi}\frac{(1 - e^{-5929/t})}{11858}. \tag{23}$$

The series expansion of the part of the expression from (23) has the form
$$\frac{(1-e^{-5929/t})t}{11858} \approx \frac{1}{2} - \frac{5929}{4t} + O(\frac{1}{t^2}).$$

Figure 3a shows the behavior of function $P\{\xi(t) = x\}$ (22) from $x$ at $t=3000$ sec, and Figure 3b shows the behavior $P\{\xi(t) = x\}$ from $t$ at $x=0.5$ in the interval $t$: 2900,…,9000. Figure 3c shows the behavior of the average value $\langle \xi(t) \rangle$ (23) from $t$ in the interval $t$: 4600,…,9000.

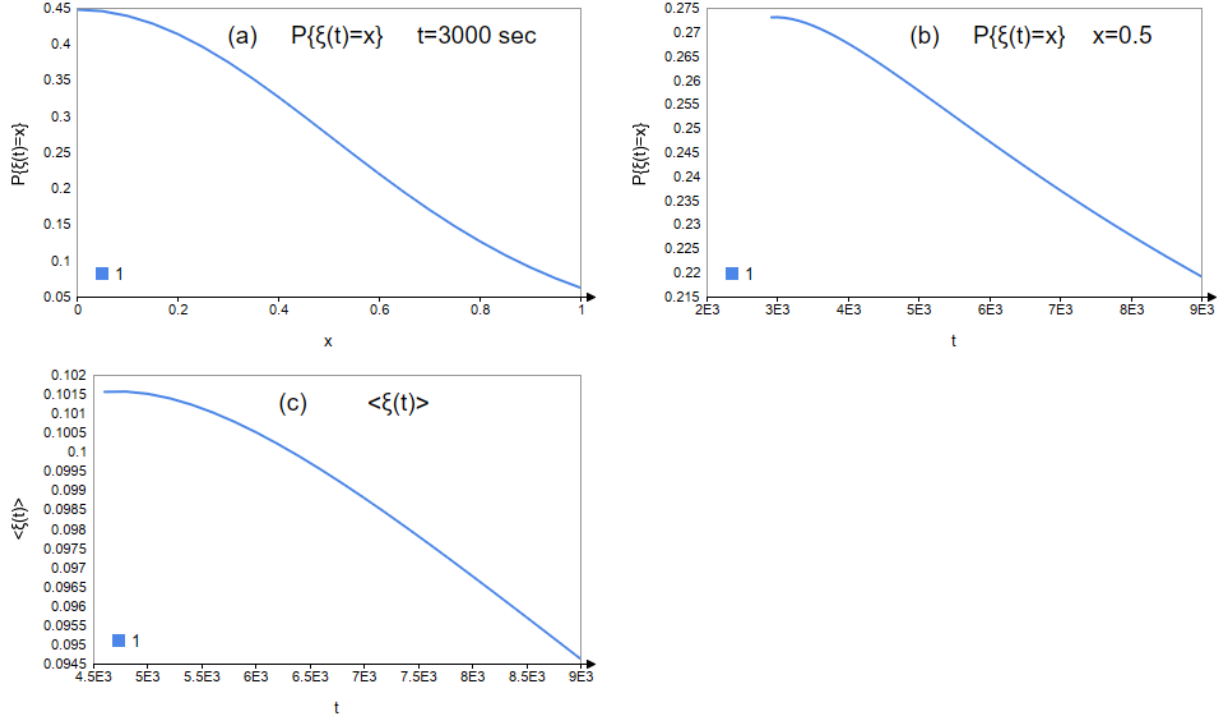

Fig. 3. Fig. 3a shows the behavior of function $P\{\xi(t) = x\}$ (21) from x=0,…,1 at $t=3000$ sec, and Fig. 3b shows the behavior $P\{\xi(t) = x\}$ from $t$ at $x=0.5$ in the interval $t$: 2900,…,9000. Fig. 3c shows the behavior of the average value $\langle \xi(t) \rangle$ (22) from $t$ in the interval $t$: 4600,…,9000.

Up to approximately $t=3000$ sec the distribution $P\{\xi(t) = x\}$ in this approximation has the form:
$$P\{\xi(t) = x\} = \frac{t}{x}\delta(t - 10x).$$

We now use the parameters $m = 6.7, n_0 = 10^8, T = 10^4, \mu = 10, \bar{\lambda} = 3.54 \cdot 10^{-5}$, $\bar{b} = 4.8 \cdot 10^{-3}$, when from (10), (17) we obtain $\rho_+(s) = 10(1 - e^{-107s})$. For such a quantity $\rho_+(s)$, the inverse Laplace transform of expression (14) also cannot be written explicitly.

If we use a series expansion in an expression of the form: (14)
$$e^{-10x(1-e^{-107s})} = e^{-10x}e^{10xe^{-107s}} = e^{-10x}(1 + 10xe^{-107s} + \frac{1}{2!}(10x)^2 e^{-2\cdot 107s} + ... + \frac{1}{n!}(10x)^n e^{-n\cdot 107s} + ...),$$

then the inverse Laplace transform is written as
$$e^{-10x}[\delta(t) + 10x\delta(t - 107) + \frac{1}{2!}(10x)^2 \delta(t - 2\cdot 107) + ... + \frac{1}{n!}(10x)^n \delta(t - n\cdot 107) + ...].$$

If, as above, we make the approximation that at $s \geq 0.05$ (which corresponds to times $t \leq 5 \cdot 10^2$ sec ) the second term in the exponent of the expression $\exp[-10x(1-e^{-10^7 s})]$ is much less than one, then in this interval the distribution density of the time of first achievement is equal to the delta function, and the distribution $P\{\xi(t) = x\}$ in this approximation has the form $P\{\xi(t) = x\} = \frac{t}{x}\delta(t-10x)$.

In the interval $s$: $0,\ldots,5 \cdot 10^{-2}$ ($t > 5 \cdot 10^2$ sec ) we approximate the dependence $\exp[-10x(1-e^{-10^7 s})]$ by the function $\exp[-65xs^{1/2}]$. The inverse Laplace transform is equal to $\frac{65x}{2\pi t^{3/2}} e^{-\frac{(65)^2 x^2}{4t}}$, the probability $P\{\xi(t) = x\}$ is given in (24), $\langle \xi(t) \rangle = \int_0^1 xP\{\xi(t) = x\}dx$ in (25);

$$P\{\xi(t) = x\} = \frac{65}{2\pi t^{1/2}} e^{-\frac{(65)^2 x^2}{4t}}, \qquad (24)$$

$$\langle \xi(t) \rangle = \int_0^1 xP\{\xi(t) = x\}dx = \frac{65}{\pi\sqrt{t}} \frac{(1-e^{-4225/(4t)})t}{4225}. \qquad (25)$$

Fig. 4a shows the behavior of function $P\{\xi(t) = x\}$ (24) from $x$ at $t$=500 sec, and Fig. 3b shows the behavior $P\{\xi(t) = x\}$ from $t$ at $x$=0.5 in the interval $t$: 500,…,9000 sec. Fig. 3c shows the behavior of the average value $\langle \xi(t) \rangle$ (25) from $t$ in the interval $t$: 800,…,9000 sec.

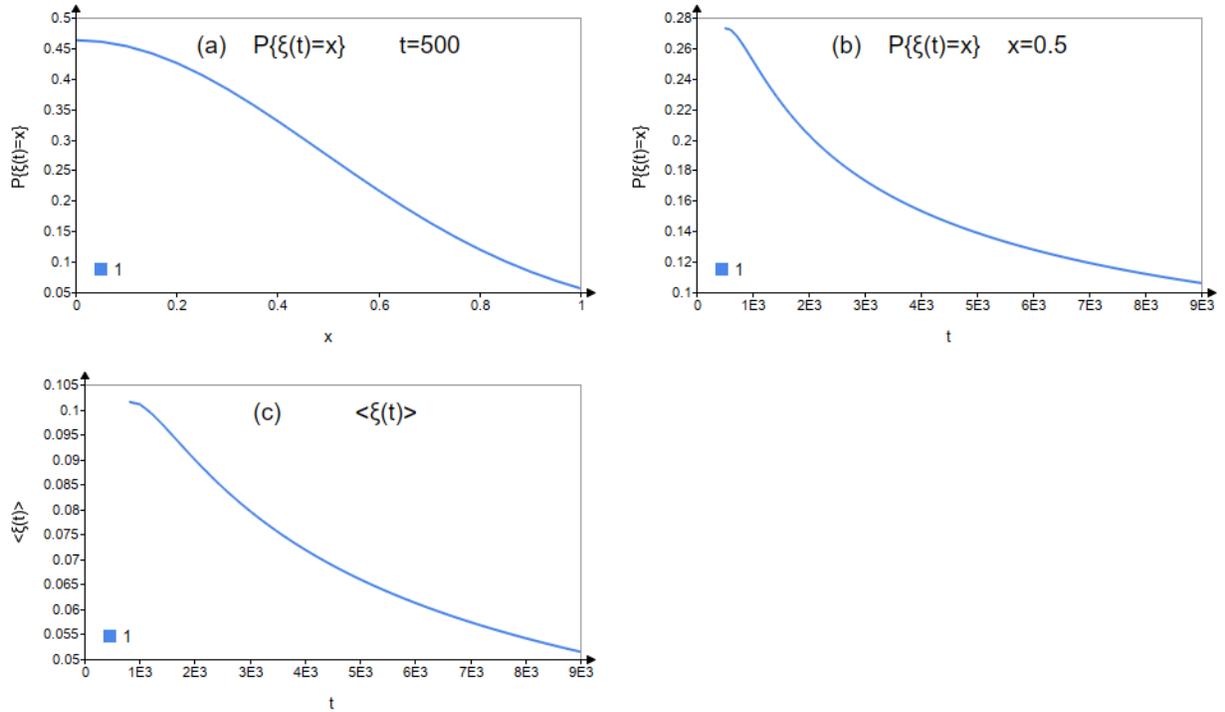

Fig. 4. The behavior of function $P\{\xi(t) = x\}$ (24) from $x$=0,…,1 at $t$=500 sec, and Fig. 3c shows the behavior $P\{\xi(t) = x\}$ from $t$ at $x$=0.5 in the interval $t$: 500,…,9000 sec. Fig. 3c shows the behavior of the average value $\langle \xi(t) \rangle$ (25) from $t$ in the interval $t$: 800,…,9000 sec.

The curves in Fig. 4 are similar to the curves in Fig. 3. But the function $\rho_+(s)$ used in the calculations in Fig. 4 and the parameter values used are close to the values that gave the correct result in the case of Fig. 2d. The similarity of the behavior of probabilities and average values of aerosol mass concentration with a large spread of parameter values (*m*=50 for Fig. 3 and *m*=6.7 for Fig. 4; also, different $\bar{\lambda}$, $\bar{b}$) reflects the universality of these patterns and the weak influence of the choice of parameters.

The boundary functional of the magnitude of the jump over a positive level was also considered. It was found that the average value of such a jump is zero. This corresponds to a non-increasing process.

The application thermodynamics of trajectories (Garrahan & Lesanovsky, 2010) and large deviation theory to the problems under study were also considered. These theories, as shown in (Ryazanov 2025b), are related to risk processes and stochastic storage processes. Although the results are qualitatively similar, the results presented in this paper are closer to experimental aerosol measurements.

Another expression can be given for the average time of stay of the process *n(t)* in the interval (*M*, *K*), *M*>*K*: $\tau_{MK}=\tau_M(1-K/M)$, where $\tau_M$ is the mean first-passage time reaching the value *M*.

## 5. Conclusion

The proposed stochastic approach is ambiguous. Thus, it is not necessary to use a stochastic storage model; other stochastic models can be used. An approximate solution to the Smoluchowski equation (6)-(8) and other approximations were used. Expressions (10) and (14) may also employ various other approximations.

One of the main goals of this article is to demonstrate the potential of applying boundary functionals to aerosol coagulation. A dependence such as 1/*n(t)* is well known (Fuchs, 1964). However, the independent approach developed in this article, based on the theory of random processes, corresponds to the results of (Fuchs, 1964) for the selected parameter values. This confirms the validity of the proposed approach. Original results were also obtained, such as the distribution of the random variable $\xi=n/n_0$. This distribution was obtained for large time periods and can be refined.

The behavior of some boundary functionals for the quantity under consideration—the mass concentration of aerosol—is trivial due to its decreasing nature. Thus, its maximum is equal to the initial value, and its minimum is zero.

A monograph (Voloschuk, 1984; Piskunov, 2010) presents a number of analytical solutions to the kinetic equation of coagulation. The results of this article do not coincide with them, but the qualitative behavior is consistent.

The choice of parameters (*T*, *m*, $\mu$,…) plays an important role. Selecting the optimal set of parameters is the subject of a separate study.

If we consider the function $k(-r)$ corresponding to the process $-n(t)$, as well as the traditional definition of the Laplace transform $\ln F(e^{-r})$, then from equation (12) we obtain an expression for $\rho_-(s) = \rho_+(s)$. In (Gusak, 2007), an expression $\boldsymbol{E}[e^{-s\tau^-(x)}, \tau^-(x) < \infty]$ for achieving negative levels is given, which coincides with expression (14) after replacing $\rho_+(s)$ with $\rho_-(s)$. In this case, the results of the article, Figs. 2-4, do not change.

In (Gusak, 2007) an expression $\boldsymbol{E}[e^{-s\tau^-(x)}, \tau^-(x) < \infty]$ is also given, which is more precise for in the case of upper-continuous processes, which includes the process $n(t)$:

$E[e^{-\gamma\tau^{-}(x)}, \tau^{-}(x) < \infty] = (1 - \rho_{-}(\gamma)/\mu)e^{\rho_{-}(\gamma)x}$, $x < 0$. However, estimates show that the contribution of the pre-exponential factor is much smaller than the contribution of the exponential factor.

Overall, the application of stochastic processes and their boundary functionals to describing the behavior of aerosol systems appears promising. The results of this article may find application in such applied problems as water purification using coagulants or the development of anticoagulants in medicine.

**References**

Abramowitz, M., & Stegun, I. A. (1972). *Handbook of mathematical functions*. New York: Dover.
Asmussen, S. (2000). *Ruin probability*. Singapore: World Scientist.
Bayewitz, M. N., Yerushalmi, J., Katz, S. & Shinnar, R. (1971). *J. Atmos. Sci.*, 31, 1604.
Borovkov, A. A. (1976). *Stochastic processes in queueing theory*. Berlin: Springer-Verlag.
Brockwell, P. J. & Chung, K. L. (1975). *J. Appl. Probability*, 12, 212.
Brockwell, P.J., Resnick, S. I. & Tweedie, R. L. (1982). *Advances in Applied Probability*, 14, 392.
Cinlar, E. & Pinsky, M. (1972). *J. Appl. Probability*, 9, 422.
Çinlar, E. (1975). *Ann. Probability*, 3, 930–950
Debry, E., Sportisse, B. & Jourdain, B. (2003). *Journal of Computational Physics*, 184(2):649.
Feller, W. (1957). *Introduction to Probability Theory and its Applications*, 2 vol. New York: Wiley.
Fuchs, N. A. (1964). *Mechanics of Aerosols*, Oxford: Pergamon Press.
Garrahan, J. P. & Lesanovsky, I. (2010). *Phys. Rev. Lett*. **104**, 160601.
Gikhman, I. I. & Skorokhod, A. V. (1969). *Introduction to the Theory of Random Processes* by W. B. Saunders Company (Courier Corporation, 1996) p.516.
Gusak (Husak), D. V. (2007). *Boundary problems for processes with independent increments in the risk theory* (in Ukrainian) (Kyiv: Proceedings of Institute of Mathematics; v. 65,) p.459.
Harrison, J. M. & Resnick, S. I. (1978). *Math. Oper. Res.*, 3, 57.
Kolodko, A. Sabelfeld, K. & Wagner, W. (1999). *Math. Comput. Simul.* 49 (1/2) 57.
Lushnikov, A. A. (1978). *Izvestiya AN SSSR. Fizika Atmosfery I Okeana*, 14, 1046.
Lushnikov, A. A., Bhatt, J. S. & Ford, I. J. (2003). *Journal of Aerosol Science*, 34, 1117.
Merkulovich, V. M. & Stepanov, A. S. (1985) *Izvestiya AN SSSR. Fizika Atmosfery I Okeana*, 21, 1064 (Russian).
Merkulovich, V. M. & Stepanov, A. S. (1991). *Izvestiya AN SSSR. Fizika Atmosfery I Okeana*, 27, 266 (Russian).
Merkulovich, V. M. & Stepanov, A. S. (1992). Izvestiya AN SSSR. Fizika Atmosfery I Okeana, 28, *752 (Russian*).
Metzler, R., Oshanin, G. & Redner, S. (eds), (2014). *First-Passage Phenomena and Their Applications*. Singapore: World Scientific.
Piskunov, V. N. (2010). *Dynamics of Aerosols*. Moscow: Fizmatlit (Russian).
Rolsky, T., Shmidly, H., Shmidt, V. & Teugel, J. (1999). *Stochastic processes for insurance and finance*. New York: John Wiley.
Rubinovitch, M. & Cohen, J. W. (1980). *J. Appl. Probab.*, 17, 218.
Ryazanov, V. V. (1991). *Journal of Aerosol Science*, 22 (Supplement 1) S59.
Ryazanov, V. V. & Shpyrko, S. G. (1995). *Journal of Aerosol Science*, 26 (Suppl.1) S643.
Ryazanov, V. V. (2011). *Applied Mathematics*, 1 (1): 1.
Ryazanov, V. V. (2025a). *Phys. Rev. E*, 111, 024115.
Ryazanov, V.V. (2025b*). Physica A: Statistical Mechanics and its Applications*, 674, 130760.
Schumann, T. (1940). *Quart. J. Roy. Meteorol. Sce.*, 66, 195.
Scott, W.T. (1967). *J. Atmos. Sci*., 24, 221.
Van Dongen, P. G. J, & Ernst, M. H. (1987). *J. Statistical Phys.*, 49, 879.
Voloschuk, V.M. (1984) *Kinetic Koagulation Theory*. Gidrometeoizdat, Leningrad (Russian)
Williams, M. M. R. (1984) *Physica A: Statistical Mechanics and its Applications,* 125, 105.
Zakusilo, O. K. (1988) *Theory of Probability and Mathematical Statistics*, **38**, 35.
Zakusilo, O. K. (1990) *Theory of Probability & Its Applications*, **34** (2), 240.